# Bosonic mode interpretation of novel STM and related experimental results, within boson-fermion modelling of HTSC

**JOHN A WILSON**

H.H. Wills Physics Laboratory,
University of Bristol,
Tyndall Avenue,
Bristol BS8 1TL. U.K.

**ABSTRACT**

This paper seeks to synthesize much recent work on the HTSC materials around the latest energy resolved scanning tunnelling microscopy (STM) results from Davis and coworkers. The STM conductance diffuse scattering results in particular are employed as point of entry to discuss bosonic modes, both of condensed and uncondensed form. The bosonic mode picture is essential to understanding an ever growing range of observations within the HTSC field. The work is expounded within the context of the site-inhomogeneous negative-*U* boson-fermion modelling long advocated by the author. This general approach is presently seeing much theoretical development, into which I have looked to couple many of the experimental advances. While this formal theory is not yet sufficiently detailed to cover adequately all the experimental complexities presented by the real cuprate systems, it is clear it affords very appreciable support to the line taken. An attempt is made throughout to clarify why and how it is that these novel circumstances and phenomena are tied so very closely to this particular set of materials.

e-mail: john.a.wilson@bristol.ac.uk

## §1. INTRODUCTION TO THE ENERGY RESOLVED STM WORK OF HOFFMAN *et al*.

Recently Hoffman *et al* (2002) in an energy resolved 4K STM study of local tunnelling conductance into Bi-2212 have uncovered that over the energy range 5 to 30 meV additional novel scattering information is forthcoming. Through a Fourier transformation of their real space local conductance maps they show that incommensurately structured elastic scattering of the incoming quasiparticles is being incurred. This scattering is quite distinct from that previously reported and ascribed to (*i*) low energy impurity effects, (*ii*) magnetic effects, (*iii*) charge stripe effects, (*iv*) vortex effects, and (*v*) the Bi-O layer superlattice modulation. The newly found modulations in the tunnelling conductance signal are much more diffuse than those from (*v*), and the action at the wavevectors involved is in additionseen to be strongly energy dependent. The dispersion characteristics manifest in the new scattering are moreover not those of the antiferromagnetic magnons of low doping, nor are they likewise those of the lattice phonons. The engaged wavevectors ($\mathbf{q}_A$ and $\mathbf{q}_B$) are found to relate directly to the basal Fermi surface geometry, $\mathbf{k}_F$, and the scattering is furthermore coupled with tip binding energies roughly equal in value to the superconducting gap. As a result Hoffman *et al* have tentatively presented their 4K STM conductance data in terms of the introduced electrons being scattered from the Bogoliubov quasiparticles. The density of states peak for the latter then should of course map out the energy below $\mathbf{k}_F$ of the bottom of the superconducting band gap as structured by the $d_{x2-y2}$ symmetry HTSC order parameter $\Delta(\mathbf{k}_F)$. However upon closer examination of the new STM results we will show that whilst at each level of sample hole doping reported the locus from the full set of conductance modulation wavevectors indeed follows a shape in *k*-space very close to the ARPES-determined normal state Fermi surface, the associated binding energies $E_q(\theta)$ consistently are significantly greater than $\Delta(\theta)$ (though less than $2\Delta(\theta)$ – prior at least to the ‘hot spot’ being reached (see below)). It will be argued that this observation is in accord with the existence of a dispersed uncondensed boson mode, as figures in the negative-*U* two-subsystem treatment of HTSC from the current author (1987-2001). Formal development in modelling of a mixed boson-fermion and/or negative-*U* form recently has been much advanced by Micnas *et al* (2002abc), by Domanski *et al* (2002abcd), and by Batle *et al* (2002,2001), Casas *et al* (2002) and Fortes *et al* (2001), with closer examination being made of questions relating in particular to the electronic specific heat, entropy and condensation energy data from Loram *et al* (1998, 2000, 2001) and to the Uemura type rendering of the μSR data (2000).

A matter that we shall return to in due course but one which at this point should be registered is the misleading reading of the true value of the superconducting gap from the ARPES spectra adopted by Hoffman *et al*. In the ARPES spectra the leading peak has to be viewed, unlike in earlier type readings of such data - and taken up now by Hoffman *et al* (2002) - as much more closely defining $2\Delta$ than it does $\Delta$. $\Delta$ actually is better monitored by the inflexion point of the leading edge. At optimal doping in YBCO-123 and BSCCO-2212 it is the maximal gap values $2\Delta_o$ that fall just short of 40 meV. Thermally activated experimentation such as specific heat and nmr, along with phonon line-width and electronic Raman analyses all concur on such a $2\Delta_o$ value ($\cong$ 320 cm$^{-1}$) [Wilson & Zahrir 1997 (§E16) and Wilson 1994 (ref 36)]. This then yields the well-known strong-coupling $2\Delta_o/kT_c$ ratio of approximately 5.5. A proper reading of $2\Delta$ directs interpretation of what Hoffman *et al*'s new results mean along significantly different lines from those followed in their paper. The correct interpretation of the ARPES spectrum from Bi-2212 has been a long and complex task – one involving the joint effects of the local boson mode and bilayer splitting, in addition to the matter of quasiparticle delocalization.

What is most striking about the new STM results, as is indicated above, is that for any given doping level ($p$') the *maximal* (*) scattering signal identified in the conductance occurs at some energy $|E_q^*|$ appreciably smaller than $2\Delta(\mathbf{k}^*,p',T = 0)$ for coupled $\mathbf{q}_A^*/\mathbf{q}_B^*$. The latter wavevectors locate a set of equivalent points within $k$-space which fall well removed from the $(\pi,0)$ saddles in the band structure, as indeed from the renowned 'hot spots' sited on lines $(0,\pi)$ to $(\pi,0)$, *etc*. at the points where those lines intersect the Fermi surface. In the spin fluctuation interpretation of HTSC the latter 'hot spots' are to be associated with the highly characteristic $(\pi,\pi)$ scattering physics (Eschrig and Norman 2002, Norman 2001, Timm *et al* 2002, Yanase 2002, Abanov *et al* 2001), pre-eminently in evidence in the striking 'resonance mode' of inelastic neutron scattering, this scattering being treated in standard nesting fashion (He *et al* 2002, Fong *et al* 1999,2000, Dai *et al* 1999, Keimer *et al* 1998). By contrast our own negative-$U$ interpretation of the $\pi,\pi$ scattering resonance involves not true *magnetic* spin excitation but a singlet-triplet spin-flip pair breaking of local pairs (Wilson & Zahrir 1997). The maximal scattering within the new STM data as well as arising in k-space as close to the $d_{x2-y2}$ nodes as to the hot points does so at a binding energy $E_q^*$ that it was noted above is only around $\frac{1}{2}.2\Delta_o(\mathbf{k}^*,p')$. It would appear the STM conductance signal becomes maximized where/when the dispersed boson mode of the postulated negative-$U$ model stands not too severely perturbed either by pair formation or by pair dissolution; within the model the latter come in the vicinity of the hot spots and of the gap nodes respectively.

The above STM scattering results are employed in the present paper as point of departure against which an extensive range of recent experimental results may be addressed within the framework of the author's long-standing mixed-valent negative-$U$ interpretation of HTSC behaviour. Section 2 presents some of the earlier detailing of the model along with a full description of the new STM scattering experiment and data. Comparison is made with recent ARPES results. Section 3 asserts the STM scattering entity to be a dispersed mode of uncondensed local pair bosons, and it contrasts the STM results with what has been recorded in inelastic neutron scattering, both of spin-flip and phononic issue. Section 4 deals conversely with the condensed boson state and with its excited plasma mode. The latter is presented as being responsible for the kinking induced in the quasiparticle bands around 60 meV, and also for the strong HREELS signal found at this energy. The very recent Nernst results are considered here and again at the beginning of Section 5. Section 5 in the main deals with the nature and consequences of the mixed-valent inhomogeneity, and turns to further new STM results. Section 6 starts once more from STM work, introducing the variety of roles LO phonons play in HTSC phenomena, in particular $\Sigma_1$ coupling to the plasma mode of the boson condensate. The consequences for phonon dispersion, Raman, IR, nmr, isotopic shift and high pressure results are all detailed. Finally Section 7 considers more closely the current theoretical state of play regarding the nature of HTSC in the cuprates. It includes the effect of a slight breakage of *e-h* symmetry in the Bogoliubov treatment, and sets this against the strong asymmetry prevalent between hole and electron carrier types within the bank of known superconductors. Shell-filling effects are contrasted with those of the band Jahn-Teller derivation as regards the sourcing of exotic non-BCS type superconductivity. Throughout indication is made of how spin-fluctuation, phonon and Marginal Fermi Liquid modellings of HTSC each fall short of what can be accomplished by pursuing the boson-fermion, resonant negative-*U*, two-subsystem approach.

## §2. RELATIONSHIP OF THE NEW STM DATA WITH $2\Delta(\theta)$ GAP DATA FROM ARPES.

Hoffman *et al* (2002) observe that the new diffuse peaks appearing in the (**r**:**k**) Fourier analysis of the STM conductance data look clearly to reflect the geometry of the Fermi surface. Specifically the two wavevectors **q** detailing the scattering events in *k*-space appear in elastic fashion at each pre-set energy *E* below the Fermi surface to span between equivalent pairs of degenerate points within the zone, with set $\mathbf{q}_A$ directed parallel to the Cu-O bonds and set $\mathbf{q}_B$ lying at $45^o$ to these. The $\mathbf{q}_A$ then span across the occupied arms of the Fermi sea, while the $\mathbf{q}_B$ span between these arms across the unoccupied parts of the zone. Note accordingly that the $\mathbf{q}_B$ lie parallel to the Bi-O superlattice modulation vector, the sharp diffraction spots for which are very visible in the STM conductance Fourier transforms [*ibid*; fig 3]. The above assignment of $\mathbf{q}_A$ and $\mathbf{q}_B$ is in full accord with the observation made by Hoffman and colleagues that as the (hole) doping is augmented so the magnitude of $\mathbf{q}_A$ decreases whilst that for the partnering $\mathbf{q}_B$ increases. Plotted in figures 4a,b of their paper is just how the measured scattering amplitudes vary with wavevector magnitude to yield the above pairings $|\mathbf{q}_A|$ and $|\mathbf{q}_B|$, and secondly how these two peak scattering values progressively change as one incrementally augments the binding energy of the STM tip. By combining this information one is enabled, upon accepting the stated dispositions for $\mathbf{q}_A$ and $\mathbf{q}_B$, to arrive at the Fermi surface geometry. The outcome actually was not displayed by Hoffman *et al*, and it is presented now in fig. 1. Included in the figure is a circle about the zone corner that for an idealized, homogeneous, band-like situation in the 2D limit would correspond to the F.S. for a hole doping *p* of (1)+ 0.16, the band filling to support maximum $T_c$ in all cuprate HTSC systems.

**{figure 1}**

The HTSC compounds are of course not ideal metals, but are very highly correlated, are not homogeneous, but chemically mixed-valent and dynamically charge- and spin-striped (Wilson & Zahrir 1997, Wilson 1998), and are not in the 2D limit, but hold measurable 3D character. In the Bi-2212 case the latter automatically will introduce direct intra-bilayer interaction and for a standard material this is going to lead to a two-sheeted $d_{x2-y2}$ symmetry Cu-O $pd\sigma^*$ band complex. Here the fuller component (re *electrons*, and the one supplying the outermost electron F.S. sheet) will be that for which the intra-bilayer interaction entails *c*-axis 'bonding' as opposed to 'antibonding' phasing between the pair of $pd\sigma^*$ wavefunctions to arise from the two Cu-O chessboard arrays per unit cell. Following much initial searching for this 3D splitting, several more recent ARPES works employing better resolution now are able to detect both components to the $pd\sigma^*$ band (Bogdanov *et al* 2001, Feng *et al* 2002a, Chuang *et al* 2001ab, Kordyuk *et* al 2002, Borisenko *et al* 2002ab), at least for optimally and overdoped material. Since the same groups, however, when working with Bi-2223 observe still only two relevant spectral features, not three (Feng *et al* 2002b), and more tellingly yet with Bi-2201 continue to find two clear features, not just one (Sato *et al* 2002, Takeuchi *et al* 2001), this justifies our continued adherence to the position recently re-expressed by Campuzano, Norman and Randeria (2002) in their very comprehensive review of the ARPES work that the universal 'peak and hump' circumstance apparent within all these spectra is primarily the consequence of strong correlation and scattering. It arises quite independently of any multisheet coupling. The consensus to emerge from these more refined ARPES determinations relating to the overall shape of the Fermi surface is that the latter departs appreciably from the circular form of the idealized surface included in figure 1 towards a cross, the latter retracted somewhat in the vicinity of the 45$^o$ nodal directions and correspondingly inflated in the saddle point regions. It is pleasing to observe now that this is just the kind of modification echoed in figure 1 by the new STM data. The curve seen there is very comparable in form to that for the fuller F.S. component in the LDA band structure calculations (Andersen *et al* 1994, Massidda *et al* 1988) and again to what is traced by the corresponding sharp peak in the ARPES work. In the ARPES data the key ubiquitous spectral features of 'peak and hump' are separated at the saddle points by ~ 90 - 140 meV, dependent upon doping level. Against this the LDA band structural work gives a bonding/antibonding (B/AB) splitting at the saddles under coherent bilayer-type interaction of up around 300 meV; in contrast the new ARPES work would support a considerably reduced value of about 85 meV. By symmetry this B/AB splitting steadily drops to zero upon moving towards the 45$^o$ direction. It is for quite different reasons that $(\pi,\pi)$ direction is the orientation adopted by the superconducting *d*-wave nodes (Wilson 2000b).

At this juncture it is well to remember that de Haas-van Alphen-type resonances have not yet been attained for the HTSC cuprates, even for that structurally most favourable of cases $YBa_2Cu_4O_8$ (Y-124) (Meeson 2002), nor in all probability will this be possible either with very highly overdoped material, witnessing the parallel null result obtained with the 3D mixed-valent *s*-electron superconducting system (Ba/K)$BiO_3$ (Goodrich *et al* 1993). All HTSC materials are characterized by chronic scattering, and that is not just because they are poorly formed or intrinsically nonstoichiometric. Indeed with $PrBa_2Cu_4O_8$, upon there compromising greatly the potential within the Cu-O basal *planes* for metallic conductivity (and indeed superconductivity), this chain-bearing material actually becomes rendered a notably improved metal overall, more coherent and Fermi-liquid-like even in its *c*-axis direction (Hussey *et al* 2002, McBrien *et al* 2002, Horii *et al* 2002). Recall with Pr-124 that, despite it being macroscopically crystallographically perfect, one finds the basal plane electric and magnetic behaviour to be totally transformed under the strong Pr*4f*-Cu*3d* hybridization. This *f-d* mixing with its spin coupling inhibits the generation of the RVB spin-singlet condition so important in allowing HTSC to arise.

Within the author's negative-$U$ two-subsystem perception of HTSC, the very severe S = 0 pair scattering experienced in HTSC YBCO-124, -123, etc. by the basal carriers near the saddle points and hot spots constitutes the very means to acquiring local pair formation via the key shell-filling negative-$U$ process. Saddle-point scattering is source not just to the superconductivity but in large measure too to the highly anomalous 'normal' state properties – and specifically to the $p$-type sign for the carriers (by reinforcement of the strong Mott-Anderson pseudo-gapping) as well as to the $T^2$ dependence to their mobility up to very high temperatures (Wilson & Zahrir 1997, Wilson & Farbod 2000). As has been elaborated upon recently by Hussey (2003), the HTSC cuprate systems exhibit large basal *e-e* scattering anisotropies, these bearing enormous $T$-independent prefactors, and the customary poor-metal *e-ph* resistance saturation no longer is pre-eminent. The above chronic saddle-point scattering is shown to add in parallel to a more normal component, seen more clearly once away from the saddles. As it so happens, and as was depicted in fig 3 of Wilson 2000b, given the existing geometries of crystal structure and Fermi surface, even the best of carriers, viz. those running in the nodal directions, are open to continued strong *e-e* scattering over into the saddle-point sinks. It hence is not surprising that the HTSC materials carry significant local character through to such high doping levels, despite standard LDA band structure analysis ascribing to the crucial $d_{x2-y2}$ $pd\sigma^*$ band a bare width in excess of 4 eV. That width reflects the high degree of $p/d$ mixing experienced in these materials as the copper $d$-states drop down towards shell closure through the oxygen based $p$-states (Wilson 1972). The above extremely strong *e-e* scattering persists to out beyond $p_h$ = 0.3 (Takagi *et al* 1992, Kubo & Manako 1992, 1994, Nakamae *et al* 2002), *i.e.* to well beyond where the novel superconductivity is any longer sustainable.

**{figure 2}**

In figure 1 the tip energy has been incoporated as running parameter. Indicated as well is the point at which the STM scattering signal is greatest. The latter develops well away both from the saddle point and the 45° nodal directions. In fact the present scattering drops below detection considerably in advance of reaching either such limiting position. With figure 2 the next step now is presented of displaying the variation of the above $E_q$ as a function of angular position $\theta$ around the Fermi surface. $\theta$ here is measured about the zone centre anticlockwise from the $\pi$,0 saddle direction. This plot permits direct comparison to be made between energies $E_q(\theta)$ and $2\Delta(\theta)$, the latter as given approximately by the leading sharp peak formed below $T_c$ in the ARPES spectra. Angular peak position plots have been published by Mesot *et al* (1999ab) for a variety of under-, optimally and over-doped BSCCO-2212 samples. It is this peak data that we incorporate here in figure 2, although reading it now as being much more closely related to $2\Delta$ than to $\Delta$, unlike originally taken by Mesot and coworkers. Both the $E_q(\theta)$ and the $2\Delta(\theta)$ values prove somewhat difficult to specify exactly and the plots call for a certain amount of smoothing of the raw data. Nonetheless several features stand out from the results displayed in figure 2: (*i*) $E_q(\theta)$ is effectively a linear function, (*ii*) $2\Delta(\theta)$ is, as was emphasized by Mesot *et al*, as extracted far from being pure $d_{x2-y2}$ in form, (*iii*) despite the $2\Delta(\theta)$ plot bowing towards diminished $|2\Delta|$ versus the idealized sin 2$\theta$ form, it ultimately does turn toward zero as $\theta \rightarrow 45°$, (*iv*) $E_q(\theta)$, however, moves to

zero much in advance of $\theta = 45^o$, ($v$) the maximum value of $2\Delta(\theta)$ would appear to be attained in the vicinity of the hot spot, where the Fermi surface crosses the $(0,\pi)$-$(\pi,0)$ tie line near $\theta = 12^o$. Note no Fermi surface exists close to the $\theta = 0^o$ saddle-point direction, in particular in the 2D limit. What is viewed as being highly significant now is that across the central range of $\theta$ the STM data points $E_q(\theta)$ clearly reside (for common $\theta$) at binding energies considerably reduced as compared with $2\Delta(\theta)$, identity being gained only at the stage of maximal $2\Delta(\theta)$, reached with the hot spot location (Wilson 2000b). All such hot spot locations appertaining to the various sample underdopings would look to line up along the extrapolated $E_q(\theta)$ linear plot. Optimal doping arises considerably in advance of this extrapolation taking the hot spot location right back to the saddle axis $\theta = 0^o$. A rough examination indicates that the latter circumstance will not be encountered until $p \approx 0.3$. In all cuprate HTSC systems such a doping level in fact is where the last traces of superconductivity are recorded. Within the modelling we follow this extinction occurs because by $p \approx 0.3$ there no longer is retained any well-defined two-subsystem (mixed-valent) character: the materials have been brought to a more standard Fermi liquid condition and our high correlation negative-$U$ HTSC scenario rendered ineffective.

## §3. THE LOCAL PAIR UNCONDENSED BOSON MODE AS SCATTERING OBJECT.

Norman and coworkers have for some time advocated that the peculiar form of the ARPES spectrum below $T_c$, with its peak, dip, hump structure, points to the interaction of the quasiparticle states with some bosonic mode (Norman and coworkers 1998ab, 2001a). The natural implication of the presence of such a feature is that it relates critically to the superconductivity if not actually causing it. Not only does the above spectral activity figure at the gap antinodes, but at the gap nodes too there is evidence of a perturbation of simple behaviour, a clear kink being evident in the quasiparticle dispersion curves, with this growing more marked below $T_c$. When Johnson *et al* (2001) first reported the latter behaviour they attributed it to spin fluctuations. Subsequently it was suggested by Lanzara and coworkers (2001), from analogy with what is found in a standard strong-coupling superconductor like Mo (Valla *et al* 1999), that phonons were responsible. The energy however is now too large (~ 60 meV) to be appropriate for phonons in general, and the situation would have to be more akin to some specific optic phonon coupling, such as is present in a CDW/PSD system, *e.g.* 2H-$TaSe_2$ (Valla *et al* 2000). Although this action might relate in the cuprates to the incipient stripe formation, it is then not evident why the kinking would increase sharply at $T_c$ to present a component growing in magnitude as the superconducting order parameter. Norman and colleagues would, despite spin-singlet pseudogapping being strongly in evidence (Fong *et al* 1999, 2000, He *et al* 2002, Keimer *et al* 1998, Dai *et al* 1999) now appear to have settled for a magnetic attribution to the mode's origin (Eschrig & Norman 2000, 2002a), largely in view of the similarity in energy to the much discussed 'magnetic resonance' excited at **Q** = $(\pi,\pi)$ through inelastic neutron spin scattering – for optimally-doped Bi-2212 a quite sharp feature positioned around 43 meV (Fong *et al* 1999). The alternative perspective advocated already in Wilson 2001, 2000b is that the mode being sensed in the ARPES experiments is associated with local bosonic pairs – pairs arising via the shell-closure, double-loading charge fluctuation negative-$U$ process. The latter proceeds in line with the energetics set out in Wilson 1988/1972, and substantiated, it is claimed in Wilson 2000a by the laser pump-probe experiments of Stevens *et al* (1996) and the thermomodulation spectroscopy of Holcomb *et al* (1997). This interpretation subsequently received added support in similar, more wide-ranging optical work from Little *et al* (1999, 2000), Kabanov, Demsar and coworkers (1999a,b, 2000, 2001) and above all Li *et al* (2002), results that were examined at length in Wilson 2001. The situation throughout has been perceived as a fragmented two-subsystem one (see fig 4 in Wilson 1988), often at low doping driven dynamically towards stripes (Wilson & Zahrir 1997, Wilson 1998). This highly perturbed geometrical condition (not entirely unlike that in a quasicrystal) assures, when taken in conjunction with a negative-$U$ pair resonance stationed in close proximity with the Fermi energy, that the level of particle-particle scattering will be extreme. A high flux of interchange between fermionic and bosonic states, as well as between boson states which are condensed and those that remain outside the Bose condensate, is inevitable.

Calculations of a negative-$U$ Hubbard type consistently have indicated that when, as here, the overall *effective* negative-$U$ value (per pair) emerges as being of the order of the bandwidth, the greatest $T_c$ values will at that stage be met with (Gyorffy *et al* 1991, Litak & Gyorffy 2000, Chen, Levin & Kosztin 2001, Kopec 2002). In the present case $|U_{eff}|$ indeed is like $W(d_{x2-y2}) \approx 3$ eV (Wilson, 2001, 2000ab). As pursued by Micnas *et al* (2002abc), Domanski *et al* (2002abcd) and Batle *et al* (2002, 2001), theoretical (homogeneous) negative-$U$ boson-fermion equilibrium models recently have been much extended beyond the original work of Micnas, Ranninger and Robaszkiewicz (1990) and of Friedberg and Lee (1989). A key matter specifically to be addressed now by Casas, de Llano and coworkers (2002) is that of boson pairs possessing non-zero center-of-momentum, in conjunction with individual bosons instantaneously remaining unincorporated into the Bose condensation. The latter pairs will, unlike the condensate itself, of course show considerable dispersion under the electronic/thermal coupling. A sizeable population of uncondensed bosons is able to develop in step with the number of condensed bosons (both negative-$U$ pairs and seeded Cooper pairs). Such uncondensed bosons have been suggested earlier (see Wilson 2001) as being likely source of the extra component of a.c. conductivity discovered to present itself in roughly order parameter form below $T_c$ within the 100 GHz time domain transmission spectroscopy data from Corson *et al* (1999, 2000). As the system becomes more underdoped the *fractional* content of these uncondensed pairs grows as the negative-$U$ state itself drops away from $E_F$ to higher binding energy. This parallels growth in the $2\Delta^*$ value witnessed as a pseudogap.

In developing their bosonic mode modelling of the ARPES data a considerable advance recently has been made by Eschrig and Norman (2000, 2002a) in properly incorporating into the self-energy analysis of the mode/quasiparticle interaction the effects of the multi-layer coupling. There emerges a clear picture of just how the peak, dip, hump structure develops within both the bonding and the antibonding components to the ARPES signal. It is the uppermost (*i.e.* antibonding) component that it is revealed bears the dominant peak marking fairly closely the maximal value of $2\Delta(\theta)$ – just 32 meV in the underdoped ($T_c$ = 65K) BSCCO-2212 sample examined by Eschrig and Norman. With optimally doped material the above energy has risen to 38 meV. Note the latter binding energy still stands some way short though of the 43 meV of the 'resonance mode' so to characterize the polarized inelastic neutron spin-flip scattering (Fong *et al* 1999). The latter resonance I earlier have associated with complete destruction of a local pair boson. Such identification matches the above numbers, the 38 meV per pair being allotted to an S = 0 separation of the pair of electrons into individual fermions, with then the additional 5 meV imparted in the spin-flip to one partner, this marking the minimum spin gap magnitude. The individual components of the pair find themselves now back near the hot spots on the saddles from which they were abstracted in the original negative-*U* pair production process. The above pair elimination calls for a net momentum transfer of $(\pi,\pi)$ to the electrons drawn from each inelastically scattered neutron, with $(\pi - \varepsilon,\varepsilon)$ going to the one emerging on the *x*-axis saddle and the balance of $(\varepsilon,\pi - \varepsilon)$ passing to the one moving to the *y*-axis saddle (see Wilson 2002b, fig 3). For the case of bilayered YBCO and BSCCO what is noteworthy in the neutron scattering process is that it occurs associated in the main with a large incommensurate *z*-axis momentum transfer – one in direct space which corresponds to the Cu-O bilayer spacing. Eschrig and Norman (2002) following the lead of Fong *et al* (1999, 2000) and Bourges *et al* (1997) interpret the latter detail as evidence of the dominant magnetic aspect to conditions prevailing in the material and regard the bonding and antibonding channels evident in these bilayer systems as associated respectively with 'even' and 'odd' parity magnetic spin coupling. They pursue this direction despite the sizeable spin gap and without any evidence in optimally or overdoped material of a genuinely magnetic correlation length divergence towards low *T* (Bourges *et al* 1997, Zavidonov & Brinkmann 2001, Mook *et al* 2002). By contrast the negative-*U* point of view would see the *z*-axis momentum transfer as being required by charge symmetry as a local pair is undone and its components separated into nearest-neighbour cells. The two electrons, being left spin-parallel by the action of the neutron, no longer are permitted to reside within the same Cu-O coordination unit (as $d^{10}p^6$). The process is illustrated in detail in figure 3.

It has to be pointed out here that the above *z*-axis 'complication' is in no way essential to the viability of the HTSC mechanism, it relating solely to pair break up. It simply is a symmetry requirement imposed by the crystal structure. When, as with Tl-2201, one has a single layer structure the neutron spin-flip scattering peak centres on the $k_z$ = 0 plane (He *et al* 2002). It remains to be observed how similarly this crystal structure will simplify details in the ARPES spectrum of Tl-2201.

**{figure 3}**

The $\{\pi,\pi\}$-point composite bosons are able by virtue of their negative-*U* standing (within the high-valent subsystem) to take up an energy location that is near-degenerate with the Fermi energy. Through elastic boson-boson collisions these bosons are from this state capable of being rendered condensed **k** = 0 objects of comparable energy. These are open in turn to becoming seeding agents towards Cooper pair formation for further Fermi surface quasiparticles in more standard +**k**/-**k** fashion. The addition of these Cooper pairs will drive up somewhat the system-wide superconductive coherence length. The negative-*U* bosons in the above way may potentially introduce an overall level of pairing wherein the triggering zone-edge component may well stand numerically the minority species.

In addition to the two types of condensed composite negative-*U* boson, to the derivative Cooper pairs, and to any remaining unpaired fermions (in particular present in our current d-wave situation), one must as well anticipate, as was suggested earlier, heavy undisrupted bosons that have been excited from their condensate through phonon interaction, clearly also in play in HTSC processes. Inelastic neutron scattering experiments for example bring to light strong modification in the dispersion of certain basal longitudinal optic phonon modes at very short wavelength ($|\mathbf{k}| \geq \frac{1}{2}(0,\pi)$) (Reichardt 1996, McQueeney *et al* 2001, Pintschovius *et al* 2002). These particular phonons possess the momentum necessary to transfer a composite boson into the saddle regions of the (closely degenerate) Fermi surface as a thermally/electronically excited bosonic quasiparticle, outside either the $\mathbf{k} = 0$ or $\pi,\pi$ condensate. Such excited bosons will be of appreciable lifetime since holding in general binding energies somewhat smaller than for the Bogoliubovon pairs at identical positions in *k*-space. Wherever the dispersed excited bosons would either drop below the chemical potential in the superconducting state (as near the saddles) or rise above the Fermi energy for the unpaired quasiparticles (as near the nodal *k*,*k* axis), there the excited composite particles will become unstable and disintegrate. In the above way we may understand how it is possible to find materialize a bosonic mode having many of the attributes displayed by the novel STM scattering results introduced in §2. The detected mode departs upwards from the $2\Delta(\theta)$ gap energies in an ever-increasing fashion as one shifts away from the hot spot location toward larger $\theta$. While the condensate coupling retains a finite $2\Delta(\theta)$ binding right around to the nodes, the excited bosons disintegrate considerably in advance of reaching the $\theta = 45^{\circ}$ orientation. The excited boson mode stands sharpest at mid-range $\theta$ values because there the excited state lifetime is longest, least affected by the high electronic activity proceeding around both the saddles and the nodes.

## §4. THE CONDENSED BOSON MODE.

It has at this point to be emphasized that the above dispersed mode is distinct from the bosonic mode concentrated upon by Eschrig and Norman in their ARPES-related work (2002ab, 2000). The latter mode runs virtually undispersed throughout the outer part of the zone. In the present work that mode will be associated directly with the condensate itself of local pair bosons and lies to higher binding immediately below the dispersed mode of *un*condensed pairs. Extrapolation of the line relating to the dispersed mode in figure 2 down into the $(\pi,0)$ saddle-point region serves to place the ground state 'bosonic resonance' (labelled $\Omega_{res}$ by Eschrig and Norman) at approaching 60 meV, *i.e.* for this *under*doped sample sited at a somewhat greater binding than its maximal superconductive gap energy per pair, $2\Delta_o$. Conversely for the *over*doped ARPES sample dealt with by Eschrig and Norman (2002) their self-energy analysis would indicate the little dispersed mode in that case to sit just *above* the somewhat diminished $2\Delta_o$ level operative there. The more underdoped (*i.e.* ionic) the system is, the farther below the Fermi level the local pair negative-*U* state resides. We shall return later to this second mode and to its relation to the neutron scattering resonance peak.

With slightly underdoped 2212-BSCCO it has been observed that 60 meV is an energy in fact echoed in the self-energy driven 'kink' feature within the ARPES-determined quasiparticle dispersion curves, and readily visible once clear of the saddle direction (Johnson *et al* 2001). There, under the prevailing crystalline and condensate symmetries, the superconductive and bilayer gapping effects become diminished and the state structure is simpler. The persistence of this observed band kinking to well above $T_c$ signals the continued presence there of minority negative-*U* local pairs, whilst the loss with $T_c$ of the Meissner effect, *etc*. registers the extinction at that point of the majority Cooper pair population, and with it a relinquishing of global phase coherence. Note in the present view the spin pseudogap existing to a hundred degrees and more above $T_c$ is attributed not so much to superconductive inter-pair phase fluctuation as to dynamic RVB spin gapping. As stated in Wilson (1988,1997) this RVB gapping is perceived as the means toward preparing the system for the production of local-pair spin-singlets, in addition to being key to their preservation. The absence in the c-axis IR conductance of any anomalous peaking just above the superconducting energy gap within underdoped yet 'well-formed' Y-124 corroborates that the fluctuational behaviour in evidence far beyond $T_c$ cannot be primarily superconductive in character (Tajima *et al* 1997, Ioffe & Millis 2000) . The extensive participation of the bulk of fermions in the seeded Cooper pairing is demanded before these systems can take up an overall condition rendered somewhat more standard in its transport properties. Once below $T_c$ the dramatic fall-off in the chronic saddle-point scattering of local pair creation and excitation brings very rapid growth in the residual electronic mean free path, as is revealed for example in the thermal Hall data (Krishana *et al* 1995) and in the very steep decline in the nmr relaxation rate (Takigawa & Mitzi 1994).

Just what the nature of the low dispersion $\Omega_{res}$ mode is, and of its relationship to the neutron scattering peak, has been most hotly debated (Kee *et al* 2002, Abanov *et al* 2002b). The uniqueness of both matters within the superconducting field suggests a very close link with the local pair condensate itself. This likewise is the case as regards the observed strong and very specific coupling into the problem of the basal-plane, Cu-O bond-stretching, longitudinal optic phonons. Over the outer parts of the zone these particular phonons find themselves severely depressed in energy (Reichardt 1996, McQueeney *et al* 2001, Pintschovius *et al* 2002, Reznik *et al* 2002), and in a manner quite dissimilar (see §6) to that in some regular Fermi surface dictated soft-mode CDW/PLD circumstance (Moncton *et al* 1977). The fact that this coupling, as with the kink in the quasiparticle dispersion above, is observed too in the *k*,*k*,0 direction rules out any connection with stripe phase formation. The virtually dispersionless form to the ARPES-revealed mode as contrasted with the mode detected in the new STM work is striking. It brings to mind the phenomenon of second sound in liquid $^4$He. Upon looking into the relevant literature it is quickly uncovered that the appropriate excitations have long been postulated for strong-coupling superconductors. Furthermore, their study was in actually extended some years ago to 2D geometry by Belkhir and Randeria (1994) in the wake of developments with the cuprates – indeed it was performed within a negative-$U$ context.

What was explored by Belkhir and Randeria is how the collective mode spectrum of a (homogeneous) superconducting system can evolve as one progresses from BCS weak coupling towards the hard-core boson régime across the intermediary crossover régime wherein the cuprates clearly reside. That paper would indicate the existence of a smooth linkage between the Anderson mode established in weakly coupled BCS superconductors and the Bogoliubov sound mode applicable to the extreme local pair limit. The authors adopt a site-*homo*geneous negative-$U$ Hamiltonian and then proceed within a generalized RPA formalism to examine the crossover régime. They turn specifically to a 2D geometry and in addition make the important incorporation of screened and unscreened Coulombic interactions, highly relevant for short-range quasiparticle pairing. The outcome is that the collective mode excitations in the crossover region exhibit small but finite dispersion as $\sqrt{q}$. The energy of the mode, while being raised somewhat above the hard-core boson plasma excitation (which would be rather low-lying being governed by $1/m_b \propto t^2/U$), displays very considerable departure from the high-lying Goldstone-like zero-dispersion behaviour of the Anderson mode, the result of the increased mass of the bound pairs within the strong-coupling lattice-type modelling. For conditions where $T_c$ maximizes (*i.e.* where $|U|/W \approx 1$; see Micnas 2002abc, Gyorffy *et al* 1991), the mode is able to take on a dispersion that amounts to a not insignificant fraction of $W$. Simultaneously it is augmented in absolute binding energy to sit well below the Fermi velocity-dictated plasma energy of the Anderson mode for weak coupling. Under the real *in*homogeneously doped circumstances of the HTSC cuprates one envisages the above plasma frequency as becoming depressed not simply because of the raised value of the effective mass for the individual bosons but in addition by virtue of their relatively low and spatially variable population density.

In probing the dielectric/optical response of a material it is well known that a plasma excitation is best monitored through the electron energy loss (EELS) function $\mathrm{Im}\{-1/(1+\varepsilon(\omega,\mathbf{q}))\}$. It is this same function that one looks to also for information regarding longitudinal optic phonons. HREELS work on a metal demands high surface quality, in effect it monitoring the surface resistivity. The low temperature cleavage of BSCCO affords one an excellent opportunity to secure reliable plasma excitation results, a project in fact carried through back in 1992 by Li, Huang and Lieber (1992,1993). Sure enough what was reported was a single strong spectral peak centred at 60 meV, this displaying, what is more, a thermal behaviour that clearly associates it with the superconductivity. The above workers actually related the signal to pair breaking. Notwithstanding this they reported the 60 meV feature to bear a $T$ dependence which despite looking to collapse sharply at $T_c$ would if treated as a mean-field BCS response actually extrapolate to an ‘onset’ temperature of 150 K. Accordingly it would seem that our boson mode experiences appreciable energy loss only as it becomes heavily damped upon the engagement once below $T_c$ of the majority Cooper pair population and global coherence.

The presence of a multi-sourced pseudogapping up to temperatures of 150 K and beyond is manifest in a great many types of experiment, not least optical, but the cleanest indicator to date that this temperature range actually retains some 'superconductive' content is provided by the very recent Nernst effect measurements from Y. Wang and colleagues (2003). The Nernst effect involves the production of an electric cross-potential (y-axis) signal as the passage of charge down a (x-axis) thermal gradient is subjected to a large and mutually perpendicular (z-axis) magnetic field – the thermal equivalent of the Hall effect. The signal issues from normal carriers besides entities associated with the superconductivity, but of course whenever the latter exist they are dominant. The above paper in fact focusses on the signal which can come from the flux vortices once these enter their 'liquid state', depinned from the lattice and able to drift in the thermal gradient. However because the signal Wang *et al* record clearly extends up 20 or more degrees above $T_c$ it is evident a local pair boson attribution becomes most appropriate.

## §5. EXPRESSIONS OF THE MIXED-VALENT INHOMOGENEITY.

What in addition is so instructive about these Nernst data from Wang *et al* (2003) is that they supply a direct means of assessing $H_{c2}$ (at which applied field the strong Nernst signal component is naturally taken to zero). Although with most HTSC samples $H_{c2}$ lies well beyond the 30 teslas available experimentally to Wang *et al*, it was discovered, in consequence of there emerging simple scaling rules both for reduced fields and temperatures, that the zero temperature critical field can be extracted as a function of (under-)doping right up to 150 tesla (for $p = 0.08$). The $H_{c2}(0)$ values so deduced are directly convertible to the corresponding coherence lengths $\xi_o$ $\{= (\phi_o/2\pi H_{c2})^{1/2}\}$ and confirm the latter as a monotonically rising function of doping $p$ (i.e. of metallicity), $\xi_o$ being just 10 Å down at $p \sim 0.05$ but close to 20 Å (or $5a_o$) by optimal doping. This reflects that as the participation of Cooper pairs is advanced the extreme local character of the superconductivity becomes relaxed somewhat. $n_s$ grows in proportion to $p$ through this doping range (Uemura 2000), though the pairing force itself ultimately peaks, and so accordingly then does $T_c$. For some time it has been established via analysis of the electronic specific heat following Loram and coworkers (1998, 2000, 2001) that the actual condensation energy per dopant charge is strongly diminished to either side of a 'critical doping' (just above that for optimal $T_c$), especially towards the underdoped side. As was pointed out in Wilson (1988), concentrations of hole doping of the parent Mott insulator very close to this optimal level are facilitated in their optimization of properties by virtue of the percolation limit which they mark within the two-subsystem basal plane geometry.

In view of the above it is rather strange Loram, Tallon and Liang in a recent preprint (2002) have chosen to play down the fragmented circumstances prevailing in the cuprate systems, particularly when underdoped. They declare that the rather sharp aspect to what occurs at critical doping would imply a more uniform state, and they advance their view by reference to sharp signals seen in nmr work, notably for $^{81}$Y and $^{17}$O. However the type of disorder being addressed in the above paper is restricted to that associated solely with static doping disorder and frozen charge segregation, as the recent STM results could at first sight appear to flag (Pan *et al* 2001).

In all HTSC systems (bar $YBa_2Cu_4O_8$ and fully oxygen-loaded $YBa_2Cu_3O_7$) there automatically exists the substitutional or interstitial disorder entailed to raise the carrier dopant count to the desired level. At low temperatures this atomic substituent disorder *is* frozen and generally random. However the key question as regards the metallic and superconducting properties (Y-124 included) is what is the spatial distribution of the coupled hole content (beyond unity per Cu). What is recorded becomes then a matter of time scale for the particular experimental probe employed. Wherever charge carrier dwell times are long compared with the characteristic probe time one will obtain a broad and multi-sited signal, whereas, and specifically in nmr, if local dwell times are (relatively) short some motional narrowing of the signal can ensue. It has been demonstrated in recent zero-field $^{63}$Cu nmr work (from LSCO) (Singer *et al* 2003ab) just how upon thorough analysis the spectra gathered clearly uphold a situation wherein fluctuating regions (if not domain walls) of segregated charge indeed arise. It would be most valuable to have this work repeated now with high quality Hg-1201, a system where the hole doping is procured via interstitial oxygen rather than cationic substitution. The mercury materials, as with all others examined, certainly provide evidence of a $^1/_8$ anomaly (Farbod *et al* 2000, Wilson & Farbod 2000). Within the fluctuational mêlée the effective doping level in the LSCO case has proven to be such that it is able to stand locally far removed from the mean value (Singer 2003ab), to such a degree that magnetic moments in fact can emerge at select sites even within mildly underdoped material. This is what was argued in my paper of 1998 leads to the weak spin-flip effects recorded in μSR (Watanabe *et al* 1992, Sonier *et al* 2003) and neutron diffraction (Mason *et al* 1993). Indeed it is the Swiss cheese view of the situation, at one time supported by Radcliffe, Loram and coworkers (1996), although now given a dynamic time scale – more like cheese fondue. The percolation threshold is a sharp threshold and clearly it is one of considerable import for global superconductivity within these strongly coupled systems. The μSR results revealed early on how the averaged $n_s$ count achieved under hole doping $p$ climbs linearly, and with $n_s$ $T_c$ likewise, as monitored in transport related properties – for the case of μSR via the screening controlled penetration depth (Uemura 2000). Under thorough analysis the specific heat results (Loram *et al* 1998, 2001) bear witness, however, that a fully proportionate condensation energy is not achieved in advance of critical doping; only with the latter does proper superconductive coherence between the different entities and structural fragments within the system become fully attained. It is in this way that when underdoped the HTSC systems come to acquire glassy characteristics below $T_c$ (Müller *et al* 1987, Deutscher & Müller 1987).

A striking close up affirmation of the local conditions prevailing in the superconducting state of HTSC materials is to be found in the low temperature, high spatial and energy resolution STM mappings of BSCCO samples published in a fairly recent paper from Lang, Davis and coworkers (2002). Let us inspect this remarkable new data from the perspective of the current paper. Spatially one observes there a condition highly fragmented as regards energy gap magnitude, and this within a sample of bulk crystallographic near-perfection. The latter categorization neglects (*i*) the dopant excess oxygen, (*ii*) some Bi on Sr sites, and (*iii*) the Bi-O layer supermodulation. These atomic 'imperfections' all are frozen in at 4.2 K and their electrostatic potential creates a time-invariant backdrop for the dopant charge. The latter charge adjusts and fluctuates at the unit cell level to establish the dynamic two-subsystem environment wherein the boson-fermion negative-*U* superconductive action proceeds. In the above energy-resolved tunnelling work one observes some nano-regions to settle into gaps of around 39 meV, with others up around 58 meV and a bridge of values between the two. Lang *et al* term the former locations $\alpha$-regions and the latter high gap regions $\beta$-regions. From what we have argued already we assert that the above $\alpha$ and $\beta$ regions are to be associated with Cooper pair and with local pair predominance respectively. In line with this identification Lang *et al* record the following: (*i*) as hole doping progresses the $\beta$-regions diminish in relative *areal* weighting – in a lightly *over*doped sample they form just 10% of the field of view as compared with 50% in an underdoped sample of $T_c$ = 79 K; (*ii*) the conductance within the majority carrier, small gap $\alpha$-regions is appreciably better than in the local pair $\beta$-regions, (*iii*) the local conductance signal amplitude and the coupled maximal gap size accordingly convert in *anti*phase upon STM tip transference from the one type of region to the other; (*iv*) only the low gap (Cooper pair) $\alpha$-regions display an impurity ($\leq$ ½% Ni) intragap resonance signal under a *positive* applied voltage (+18 meV), as would come with Bogoliubovon-like hole state behaviour. While these results are generally supportive of our present modelling one aspect they cannot by their nature show – namely any appreciable organization away from random towards striped geometry given the transient form to the latter organization under charge hopping, as evidenced in the Cu nmr work (Singer *et al* 2003ab). The intermediate gaps provide some register of this activity. The $\alpha$ and $\beta$ regions are though remarkably sharp-edged, and are remarkably consistent in their interior gap magnitudes – and would justify in this our hitherto employed 'two-subsystem' terminology.

The above discussion of the mixed-valent inhomogeneity and of its consequences for these systems is very much in line with 'simple' theoretical modelling already attempted within the context of a negative-*U* Hubbard Hamiltonian. Working with *s*-wave coupling symmetry and adopting the Bogoliubov-de Gennes procedure within a CPA context, Suvasini, Gyorffy and coworkers (1993) obtained interesting initial results, and the problem subsequently has been carried through in considerable detail for intermediate coupling (although neglecting Coulomb interactions) by Ghosal, Randeria and Trivedi (2001). One of the latters' main results is that for the *s*-wave case a globally effective order parameter persists, and this moreover in conjunction with a non-vanishing spectral gap. Clearly it would be of much value now to know conversely whether with strong local *d*-wave coupling a vanishing or near-vanishing spectral gap might persist, as experiments on the cuprates would in the main seem to suggest. Contrasting research made using the *t*-*J* model and with *d*-wave coupling (but adhering to the spinon-holon view of HTSC) has also made investigation of the consequences for the system of the existence of strong local disorder (Z Wang *et al* 2002) - there parametrized through the screened effective distance of the dopant ions from the active Cu-O plane.

## §6. A ROLE FOR PHONONS.

Just published this month by McElroy, Davis and coworkers (2003) is yet another paper with novel STM results to bear strongly upon the above. Following on from the 2002 Hoffman paper, they include now a reconstruction of the F.S. such as was given earlier in figure 1, though this time secured by identifying and simultaneously fitting *all* eight symmetry related elastic scattering vectors within the Brillouin zone at energy $E$, $\mathbf{q}_1$ to $\mathbf{q}_8$, apposite to the Fermi surface geometry, not just the shortest two detected previously (then labelled $\mathbf{q}_A$ and $\mathbf{q}_B$ as above). This new more detailed work however still asserts an identity between the quasiparticle Bogoliubovon states and the states registered in the STM conductance experiment. The latter by contrast we will continue to claim lie in fact at appreciably different binding energy from the former (assessed via ARPES). We judge the STM detected states do not in reality display the dispersion and intensity characteristics to match Davis and colleagues' interpretation of events. The experimental mode dispersion curve extracted in their paper remains much as that given in figure 2 above, showing if anything a slightly counter-sigmoid form, and certainly does not give one any confidence at all that it could swing round to zero binding by $\theta = 45^o$. Perhaps the best argument to support the Bogoliubovon interpretation proffered by McElroy *et al* is that fairly comparable results now are revealed to show up also at *positive* bias. We return to this point in due course.

Following on from the discussion of previous sections it is however best first to deal with the observations appended to the above new paper in its closing paragraph. The authors put on record there, along with figure 4, that the direct space conductance mapping signal, $g(\mathbf{r},\omega)$, exhibits under specific local conditions a novel fine-grained 'tweed' structure. The period of the tweed is just *one* unit cell, and it materializes within the map *only* when and wherever the tip tunnelling energy matches the locally relevant superconducting gap *maximum*. Both italicized observations serve to indicate that the tweed structure is associated with the strong scattering activity proceeding at the zone edge. Scattering under a reciprocal lattice vector G signals the occurrence of a coupled umklapp phonon process, and it is one that seemingly arises regardless of whatever the *local* gap size maximum at the saddles might actually be. In the present model such lattice involvement already has been mentioned in regard to satisfying momentum conservation when local *pairs* (*i*) are shifted between the (0,0) or $(\pi,\pi)$ negative-$U$ related states and the saddles, or (*ii*) convert at the saddles into +**k**/-**k** Cooper pairs. In both cases the net momentum per pair demanded for these changes is close to (1,0) or (0,1).

Just which phonons might be involved here is to be gleaned from inspection of the phonon dispersion curves obtained via inelastic neutron scattering. As noted already, there has been discovered to occur appreciable phonon softening (vis-à-vis the associated Mott insulating parent parent compound) of certain basal LO modes out towards the zone edge (Reichardt 1996, Pintschovius *et al* 2002). Furthermore clear intensity changes are found to develop between branches and also within a branch as a function of temperature. Even more striking yet, there at first sight would seem to be present in the spectrum 'extra' features beyond those branches anticipated from a standard normal mode phonon analysis for the given crystal structure. New extensive neutron scattering data clarifying these matters have just been obtained by Chung *et al* (2003), and the data deserve close examination now in regard to phonon coupling emanating not only from the changes above but also from the local pair boson modes discussed earlier.

Chung *et al*'s work deals in fact with $YBa_2Cu_3O_{6.95}$. Because very large crystals are required for the neutron work the samples of the orthorhombic material were not *a*/*b* detwinned. The endemic twinning has the unfortunate effect of superimposing the TO and LO phonon branches sensed by the neutrons for the given propagation and displacement directions ($I \propto (\mathbf{Q}_n.\varepsilon)^2$). Accordingly mode separation within the experimental data necessitates careful scrutiny.

**{table1}**

Basal modes associated with phonon propagation down the crystallographic *a* axis take irreducible representations designated by $\Sigma$, while those running in the *b* axis (chain) direction bear the label $\Delta$. The highest energy axial branches ${}^{LO}\Sigma_{1,4}$ and ${}^{TO}\Delta_{2,3}$ issue from a zone centre $\Gamma$ state of point group symmetry $B_{2g}$ at 72 meV, which Raman work detects too. Modes ${}^{LO}\Delta_{1,4}$ and ${}^{TO}\Sigma_{2,3}$ emerge from a $\Gamma$ state of symmetry $B_{3g}$ lying at 66 meV, and seen again in Raman spectroscopy. Already this 6 meV splitting, a consequence of the presence of the chains, is quite sizeable, and effects become more marked once well away from the zone centre. The subscripts on $\Sigma$ and $\Delta$ above specify states covered by particular irreducible representations within the axially relevant subgroups. In each axis a couple of phonon states have to be distinguished, the outcome of there being in YBCO-123 two $CuO_2$ planes per unit cell. Because inter-layer coupling in general is weak in the HTSC structures, such pairs of states find their degeneracy lifted noticeably only as there arises additional, more interesting, electronic coupling into one or other partner. In order to keep track of the situation close study of table 1 is recommended. The key action is witnessed in the longitudinal modes, as befits charge coupling, and is especially marked for one of the above ${}^{LO}\Sigma_{1/4}$ branches. These particular ${}^{LO}\Sigma$ branches have their propagation and their polarization vectors each aligned perpendicular to the chains (*i.e.* parallel to *a*), while conversely the two complementary $\Delta$ modes, ${}^{LO}\Delta_{1/4}$ , have these two vectors each set parallel to the chains (*i.e.* parallel to *b*). No odd behaviour is evident within the associated TO branches, whether in the form of any softening or splitting or intensity change as a function of temperature, and the same appears true too for the bond-bending modes at lower energy. At first sight the *apical* modes look equally uninteresting, but note that here there is present a nearly constant offset right across the zone of some 6-9 meV between the Raman and I.R. active partners, $\Sigma_1$ and $\Sigma_4$ (or $\Delta_1$ and $\Delta_4$), with the simpler symmetry $\Gamma_1$-compatible states taking the lower energy. These apical modes in spite of being formally transverse actually are dominantly bond-stretching in nature, like ${}^{LO}\Sigma_{1,4}$ and ${}^{LO}\Delta_{1,4}$. The latter two key sets of *basal* LO branches each display considerable downward dispersion into the body of the zone, but what really marks out their behaviour as being so unusual is their intensity change with wavevector, recorded in figure 5 of Chung *et al*'s paper. As against a standard shell model fitting to the phonon data, there is observed to occur an enormous gain in scattering intensity for specific branches in the vicinity of the hot spots near the zone edge, as well as mode energies becoming shifted down to ~ 55-57 meV (*ibid,* figures 5 and 6). While both the ${}^{LO}\Sigma$ and ${}^{LO}\Delta$ basal branches move into this range, the differing manner of their doing so is most revealing. What is recorded is that the effective degeneracy between the $\Sigma_1$ and $\Sigma_4$ interlayer combinations suddenly becomes lifted about halfway across the zone, and it remains so right through to ($\pi$,0). The large *k* section of the $\Sigma_1$ branch would look to have passed down in energy through the ${}^{LO}\Delta_{1,4}$ branches, which possibly remain unsplit. (N.B. there is of course no interaction here between the $\Sigma$ and $\Delta$ branches - they are in mutually perpendicular orientations within each particular twin domain.) The means whereby the depressed segment of the $\Sigma_1$ branch is strongly acquiring intensity is clearly from some further $\Sigma_1$ mode residing in the vicinity of 55 meV – and this is not the bond-bending phonon mode, which remains little changed in intensity right across the zone (*ibid*, fig. 5b). The changes wrought in the scattering intensity for the anomalous $\Sigma_1$ phonon branch segment manifest two very revealing characteristics: (*i*) as known for some time (Reichardt 1996) the effects become more pronounced with hole doping, *p*, and (*ii*) they grow with temperature reduction below $T_c$, tracking there the general form of the superconducting order parameter. Even more significant is the sudden onset to this strong interaction as the phonon wavelength is brought below $5a_o$ or 20 Å. (N.B. in the colour figures 2, 7 and 9 of Chung *et al*'s publication red denotes an ($\omega$,k) sector experiencing a gain in scattering intensity upon *cooling* down through $T_c$, whilst blue indicates an intensity loss.)

It becomes necessary to examine then what can be the nature of this $\Sigma_1$ mode to which the Cu-O bond-stretching ${}^{LO}\Sigma_1$ phonon branch is coupling so strongly. That there additionally exists clear although less dramatic coupling to ${}^{LO}\Delta_1$ and to the apical bond-stretching phonons $\Sigma_1/\Delta_1$ would implicate direct charge coupling. Our discussions of §3 immediately point to the plasma-like mode of the local pair boson condensate. The latter mode, recall, was taken to sit at binding energies numerically just smaller than for the bosonic local pair condensate ground state, and was deemed responsible for the kinking introduced near $\mathbf{k} = \mathbf{k}_F$ at binding energies of around 55 meV within the ARPES extracted (BSCCO-2212) quasi-particle dispersion curves – a feature to become more marked below $T_c$ (Johnson *et al* 2001). This energy location for the condensate plasma mode at wavevectors near the hotspot we have noted to be compatible with the ($\pi$,$\pi$)-imparting, resonant spin-flipping excitation by neutrons of individual pairs and their ejection from the $k = 0$ superconductive condensate, with ultimately the return of their components to $E_F$ at the saddles – in $YBCO_7$ an energy of 41 meV per pair. Such numbers would imply that the pair plasmon mode actually disperses slightly downwards (*i.e.* to bigger $|\omega|$) from the zone centre towards the zone edge saddle within these d-wave superconductors. (Such behaviour parallels that of the 41 meV spin resonance excitation itself, which as discussed at some length by Chubukov *et al* (2001) disperses to numerically *reduced* energies upon a reduction from $\pi$,$\pi$ in the crystal momentum being transmitted.) That comparable effects to the above arise too in the phonon dispersion behaviour of LSCO (McQueeney *et al* 2001) signals that in $YBCO_7$ they do *not* come solely as a result of the crystallographic presence there of the chains. Note the standard shell model does not indicate any such a strong divergence as that observed to occur between and within the relevant $\Sigma$ and $\Delta$ modes.

Why the ${}^{LO}\Sigma_1$ phonon mode with its propagation and displacement vectors perpendicular to the chains should couple more strongly with the plasmon mode than does ${}^{LO}\Delta_1$, with its chain-paralleling vectors, would seem largely to be a matter of compatible symmetries. The former mode involves, note, the bumping together of the chains, and this will assist carrier passage between the charge stripes, which in YBCO are known to follow the chain direction (Mook *et al* 2000). More significant is the reason why the really strong coupling sets in only as $k$ is increased beyond ~ $\frac{1}{2}(\pi/a_o)$, *i.e.* as the wavelength drops below $4a_o$. The latter distance, recall, is expressly equal to the pair coherence length $\xi$ for optimally doped HTSC cuprates, and, as pointed out by Belkhir and Randeria (1994), it is only for wavelengths inside this limit that the local pair plasmon mode is able properly to form and accordingly to couple with the lattice.

Very short wavelength charge-pair plasma oscillations are not confined to the above basal plane states. In the bi-, tri- and higher order layer HTSC compounds there is below $T_c$ at least partially coherent superconductivity in the $c$-axis direction, with pair transfer between adjoining $CuO_2$ layers ≈ 4 Å apart. These counter-cation occupied spaces are without oxygen and are somewhat more ionic than the covalent $CuO_2$ planes, making for lower local dielectric constants and damping. For tri-layer systems and above, Munzar and Cardona (2003) very recently have drawn attention to the fact that certain 'out-of-phase' interplane charge oscillations become Raman active (see their figure 1). Pointedly below $T_c$ the appropriate phonon modes are observed to acquire very large scattering intensity and often very substantial frequency modifications are met with upon passage through $T_c$ (*e.g.* in Hg-1234 the $A_{1g}$ + $B_{2g}$ $(x',x')$-polarized phonon excitation peak at 390 cm$^{-1}$ plummets by 50 cm$^{-1}$). Munzar and Cardona additionally demonstrate that the action of these interplane plasma oscillations is to generate a Raman scattering efficiency of such magnitude as clearly to be source to the anomalous form of the $A_{1g}$ *electronic* Raman signal, so long to have plagued satisfactory interpretation of this latter type of Raman work (Cardona 1999).

Comparable effects to the above may also be identified in $c$-axis I.R. results (Munzar *et al* 2001, Timusk 2003), where once below 150 K a broad peaking is observed to arise in the c-axis optical conductivity centred about 50 meV; neighbouring phonon peaks at the same time diminish in intensity. Note through all the myriad means of investigating the small energy excitation range up to 1000 cm$^{-1}$ (120 meV) a careful distinction has constantly to be made between those effects associated with the spin pseudogapping in the DOS (which is not particularly temperature dependent and sensed for example in nmr Knight shift experiments) and those lower energy effects around 400 cm$^{-1}$ (or 50meV) which develop not too far above $T_c$ and are associated with a sharp reduction there in the chronic basal scattering to afflict the fermionic quasiparticles (as sensed for example in nmr relaxation rate experiments). Both types of electronic change scale with $T_c(p)$ (McGuire *et al* 1994), but for quite different reasons. The former change relates principally to singlet pair breaking, or rather to its control through the establishment of RVB, while the latter change relates to pair creation, parallelling the rise in hybrid coupling between the fermionic quasiparticles and negative-$U$ bosons.

Because the pair plasma mode is interacting so directly with the lattice vibrations, and strongly once below $T_c$, it is not surprising that oxygen isotope effects have been extensively recorded for HTSC systems (Keller 2003). The latter effects however are quite different in form from those associated with the customary, phonon mediated, retarded interaction of the BCS mechanism. Indeed it is discovered under *p* change within any given HTSC system that the isotopic shift in regard to $T_c$ itself pointedly is dropping to zero at the very composition to support the highest $T_c$ – or, more probably, the highest condensation energy per pair, namely the critical doping $p \approx 0.19$. Very significantly there is in operation however at precisely this juncture a strong isotopic mass exponent below $T_c$ in regard to the penetration depth $\lambda$ ($\propto n_s/m^*_b$). Since $n_s$ is not appreciably affected by isotopic substitution this must signal an isotopic shift to be operative deriving from modification to the bosonic effective mass, $m^*_b$. The *electronic* masses look to be responding here, via the plasma mode coupling discussed above, to the eigen-energy changes imposed isotopically upon the phonon modes. When cooling from 300 to 4 K, $m^*_e$ itself has been determined to mount steadily from 2 to 4 $m_e$ (Munzar *et al* 2001, Timusk 2003).

Because the kinking in the quasiparticle dispersion curves and likewise the development of the peak, dip, hump structuring to the ARPES EDCs each become more pronounced with increase in the layering number, *n*, across a sequence such as Bi-2201, Bi-2212, Bi-2223 (Matsui *et al* 2003), one now would much like to know what happens to the above penetration depth isotope effect. The conjectured scenario has been that these layering effects derive from the known reduction with *n* of the *basal* plane Cu-O bond length, the latter rendering the negative-*U*, closed-shell state more established, yet without any augmentation here in pair breaking, as distinct from what happens with underdoping. It correspondingly is reported that uniaxial pressure applied *within* the basal plane brings about for underdoped material a steep growth in $T_c(p)^{max}$ (Schilling 2001), and so presumably in $n_s(T,p)$.

As is to be expected from the current modelling, the mass enhancements developing in the saddle regions should be more marked than those met with in the nodal regions of the Fermi surface. This along with a sharp growth in the renormalization of the imaginary part of the self energy below $T_c$ via the action of the boson mode is manifested very clearly in the latest ARPES data from Kim *et al* (2003) (scanning from $k_x$,0 across $\mathbf{k}_F$ along $k_y$). Apparent as well in this new data is that no fundamental difference of response exists between the bonding and antibonding interlayer coupled sub-bands. The kinking in the former sub-band is the more evident simply because it is the fuller. Above $T_c$ the antinodal coupling constant is evaluated to become rapidly much reduced (although it still stands around 1), whilst the mass (although not the scattering rate) is rendered more or less isotropic. Note the quasiparticle mass renormalization coming from the self-consistent action of the electron pair mode introduces quite different thermal characteristics to those expected if the mode were primarily phononic in nature or one of spin fluctuations.

## §7. More on the nature of the HTSC mechanism.

In the present form of modelling what in large measure dictates the sensitivity of response to parameter changes like pressure and doping is the precise location of the double-loading

negative-$U$ fluctuational state in relation to the Fermi energy – or once below $T_c$ to the chemical potential. Remember the small energy ~ -50 meV by which the state is to sit below $E_F$ is but a tiny fraction of $|U_{eff}|$, which in the present case is ≅ 3 eV (per pair) (Wilson 2000a, 2001). The latter in turn is only half the actual state adjustment energy accompanying double-loading shell closure (within the high-valent subsystem), since about 3 eV is negated in the requirement to 'back off' the Coulomb repulsion energy. Accordingly a high sensitivity of $T_c$ to all parameter changes is to be expected if these remarkable effects indeed do establish themselves as one reaches close degeneracy between the negative-$U$ state and $E_F$, and ready interconversion between charge fermions and bosons.

This degenerate condition very recently has been referred to by Domanski (2003) as a 'Feshbach resonance'. It is demonstrated formally by Domanski that when the boson and fermion subsystems are disentangled (via application of a continuous canonical transformation) the residual fermion-fermion quasiparticle interaction itself exhibits highly resonant structure above $T_c$, which for $T < T_c$ becomes reduced to that of the superconducting gap. A further key result for our purposes obtained by Domanski is that the overall boson-fermion system foregoes the particle-hole symmetry characterizing the standard Bogoliubov analysis of the simpler BCS mechanism.

Domanski is not the only one to investigate this symmetry breakage between the hole and electron Bogoliubov-related pair states within the superconducting phase, the matter having been extensively examined also by Batle *et al* (2002). One consequence of the shifting away in the Boson-Fermion model from the $2e$-$2h$ symmetric BCS limit and the mean field approximation, inherent in permitting bosonic pairings to exist and propagate independently embedded in the Fermi sea, is that uncondensed boson modes occur. These possess a novel dispersion character in that their energy varies linearly with **K**, the centre of mass momentum for the pairs. (When Cooper pairs move in vacuum, as in a simple BCS treatment, their dispersion is quadratic). The velocity for this uncondensed mode of bosonic pairs is comparable in size to the Fermi velocity of the unpaired quasiparticles. It is these linearly dispersed modes of uncondensed pairs which it has been claimed in §1-3 of the present paper are being sensed in the energy resolved STM conductance scattering experiments from Hoffman *et al* (2002) and now McElroy *et al* (2003). Because the 2D Fermi wavevector $\mathbf{k}_F$ is virtually circular about $(\pi,\pi)$, a plane incident near-perpendicularly to the energy cylinder at $\theta = 22½^o$ bearing linear dispersion from the saddle will locally intersect the cylinder in approximately a straight line, just as the STM results appearing in our figure 2 (or figure 3c of the McElroy paper) would indicate. The above mode is of appreciably greater velocity than for the plasma mode of the condensed bosons addressed subsequently in §§4-6. The velocity of the uncondensed mode being proportional to the pair coupling strength is automatically rather substantial in the HTSC materials. Because the pairing in these d-wave systems is dominated by action at the extensive saddles within the band structure, the interaction parameter $\lambda = N(E_F).V$ becomes considerable even if $V_{eff}$ itself is not enormous. Batle *et al* (2002) find indeed that $T_c$ values of 100 K may be attained with $\lambda$ values of only ~ ½ upon employing energy-shell coupling ranges limited to $h\omega_D$ (*i.e.* ~ 50 meV, for $E_F$ ~ 1 eV). These authors as a result loop their argument around in favour of phononically driven coupling, in line with the orientation adopted by Lanzara *et al* (2001). However in that case there would then be no compelling reason ever to depart from the Bogoliubov *e-h* symmetry of the traditional mechanism.

Batle *et al* (2002) cloud the situation somewhat further by noting, after Hirsch (2003), that the great majority of known superconductors have *p*-type Hall coefficients, as against most non-superconducting metals being *n*-type in their transport properties. This carrier electron-hole asymmetry is of course though far from being synonymous with the BCS related usage of those terms. The present author already has commented upon how the observed preponderance of *p*-type materials supporting elevated $T_c$ values would appear ascribable to chemical bonding effects, even within the simplest of systems. Where such effects become particularly evident is with the superconducting, high pressure, semimetallic, homopolar-bonded forms of elements like B, Si, P, S and I (Eremets *et al* 2001), and witnessed again still more recently in Li (Struzhkin *et al* 2002, Shimitsu *et al* 2002). One could also include here cluster structures like the $C_{60}$ (Wilson 1991) and $Ga_{84}$ (Hagel *et al* 2002) derivatives and the graphite-sulphur composites (Ricardo da Silva *et al* 2001). The most striking example of bonding-driven pairing outside the closed-shell effects of the cuprates and bismuthates is that of $MgB_2$, where the holes present in the $\sigma$-bonding B-B band clearly are responsible. The fact that, as with the cuprates, the lattice responds strongly to the pairing interaction – now in the unique reaction of the $E_{2g}$ bond-breathing mode – does not mean that one should invert the terminology and revert to an electron-phonon attribution. Those interested further in $MgB_2$ are directed to the work of Boeri and colleagues (2003) and of Mazin and Antropov (2003) dealing with these band Jahn-Teller type effects and with the non-adiabatic, anharmonic lattice coupling involved. Note when considering 'chemically driven' superconductivity one should take care to distinguish between the above homopolar bonded systems (of which β-HfNCl organo-solvated by $Li^+$ ($T_c$ = 25½ K) constitutes a further recent case (Wilson 1999)) and those that are reliant upon closed shells and disorder-forestalled disproportionation, such as the cuprates and bismuthates. A very recent possible example of the latter type might be $Na_xCoO_{2+\delta}$ (Wen *et al* 2003). This latter instance then would revolve around the high stability of the low-spin $t_{2g}^6$ closed subshell configuration, as intimated in § 7.2.9/10 in Wilson (1988).

Let us now return to the question of 2*e*-2*h* symmetry in the Bogoliubov sense - or, rather, the lack of it. The breaking of such symmetry is a feature of the Boson-Fermion modelling, and one may question whether it finds clear cut experimental expression anywhere. What in fact has long been known is that pair tunnelling signals in HTSC systems are not symmetric between positive and negative bias, unlike for standard BCS superconductors. Domanski and Ranninger (2002) indeed have addressed just such matters regarding tunnelling in one of a series of recent papers to employ the (site-homogeneous) B-F model. Despite formal criticism of their theoretical procedures by Alexandrov (2003), the general level of progress being encountered in this approach suggests that the technical problems pointed to here, of singular divergences and cancellations, can probably be removed by making the modelling somewhat more complex, say by incorporating the action of the spatial inhomogeneities of doping and stripes present in the real systems. The participation of uncondensed as well as condensed bosons would appear well-established from the gigahertz a.c. conductivity measurements of Corson *et al* [1999 2000; see Wilson 2001). Tan and Levin (2003) have formally advanced the matter to account now not only for the Corson results, but also those from Y.Wang *et al* (2003) regarding the Nernst effect in BSCCO, as has been suggested above in §4. They ascertain how up to temperatures appreciably higher than $T_c$ a rough mimicking of the Kosterlitz-Thouless behaviour in the former case and a persistence of the Nernst signal in the latter case strongly uphold the presence there of some local pairing. As indicated already, what fluctuates here, within underdoped systems in particular, is primarily the superconductive phase angle between poorly communicating tight pairs, though not quite for the reasons suggested by Emery, Kivelson & Zacher (1997). In this connection Tan and Levin make evident that the transverse thermoelectric coefficient relevant to the Nernst effect provides a much more sensitive probe of the local electronic pairing activity than do, say, the diamagnetic fluctuations. The latter remain not dissimilar in degree to the Aslamazov-Larkin behaviour within a standard superconductor. Ranninger and Tripodi (2002) have extended their theoretical modelling of the B-F scenario in this direction, permitting now a degree of itineracy to their hard-core bosons. They then follow the decay of coupling between the subsystems as a function of underdoping, tracking the shift away from amplitude towards phase fluctuations.

If the above discussed excited bosonic 2*e* pair modes exist for all $T < T^{\dagger}$ (where the pair fluctuation crossover temperature $T^{\dagger}$ is, note, not synonymous either with the RVB spin gap temperature or with the charge pseudogap temperature), there ought to be some positive indication of the corresponding independent 2*h* modes. It is our belief that the STM paper from McElroy *et al* (2003) could in fact contain just such an observation, a matter we flagged for subsequent comment at the end of the first paragraph in §6. They state that their + and − 14 meV scattering results are only 'roughly identical' (see figure 2H versus 2F). Because of the diffuse nature of those results it would as yet seem not feasible to affirm categorically this anticipated difference between the positive and negative bias results without considerably more care being taken. Nonetheless the general level of difference found between the ordinary $\pm V$ pair tunnelling results is encouraging. It now would be extremely interesting to try to extend the STM scattering work to higher temperatures, provided this is manageable without surface contamination.

One thing very apparent through all this work is that the reason the HTSC problem has been with us for so long is that it truly is a complex problem. The previously simplest of experiments repeatedly debouche into a remarkable pan of intricacy it may or may not be profitable to pursue. The nmr results were an early marker of this (see now Nandor *et al* 1999, Mitrovic' *et al* 2001, Singer 2003ab, and refs therein) and the ARPES data likewise have witnessed a similar transition from being treated as beguilingly straightforward to being recognized as highly complicated in nature (see Kordyuk *et al* 2003). What may be said, however, as with the evolution of this paper, is that, wherever one probes, the boson-fermion, two-subsystem, negative-*U* approach appears able to cope with more problems than it generates.

It is sufficiently complicated of itself to hold the required degree of flexibility to follow all the various twists and turns which the experimental work exposes. It is self-evident neither the phonon scenario of HTSC nor the spin-fluctuation one possess the necessary degree of built-in complexity to do this. I would refer those still thinking along the latter lines to examine the 'inversion' of the photoemission spectra recently undertaken by Verga, Knigavko and Marsiglio (2003), as they attempt to account for the observed kinking in the quasiparticle dispersion curves. The spectral form of $\alpha^2.F(\omega)$ for the mode coupling extracted there within an Eliashberg type treatment (see their figure 6) is quite unlike that which is expected for spin fluctuations, and in addition it holds far too much weight at high energies to be relevant to phonons alone.

One of the chief reasons for seeing the spin-fluctuation scenario actively supported for so long is its perceived compliance with neutron scattering data. Wherever the latter has been dominated however by data coming from LSCO (in consequence of the large crystals available) that is unfortunate because LSCO is the most ionic of all HTSC systems, deriving from its particular counter-ions (Wilson 1994). The reported RVB spin gap is there especially small (≈ 10 meV), and the spin gap temperature in fact looks to fall somewhat below the similarly small $T_c$ (Wilson 2000b). This has meant many of the observations made on cooling LSCO down below $T_c$ have been regarded as appertaining to the superconductivity when in reality they should be attributed to the RVB spin gap. Note the gap studied by Lake *et al* (1999) is dispersionless. Additionally for temperatures $T \geq T_c$ and for $E > \Delta_{sp}$ the quartet of spin scattering peaks that the latter record develop there to define quite sharply a 'spin coherence length' not dependent upon **q** and that is close to 32 Å or $8a_o$. The latter figure is of course roughly the size of the domain in the stripe structure being settled into with opening of the spin gap, as was depicted in figure 1 of (Wilson 1998). The observed cluster of four incommensurate inelastic scattering 'satellites' come specifically to decorate the *k*-space point $(\pi,\pi)$ because of the dominant 4-spin RVB square plaquet correlations. As was emphasized in Wilson (1998) the approximately $8a_o$ size of the incipient domain structure is not Fermi surface governed but is imposed simply by the numerology of the doping charge concentration.

Millis and Drew (2003) have very recently drawn attention to the seeming incompatibility between the self-energy broadened MDC peaks registered in the new high resolution photoemission work versus the measured optical conductivity. The latter conductivity is considerably better than what in customary circumstances the former widths would match, a dichotomy unexpected if events were to be covered by the spin fluctuation scenario. However, as we have seen, the strong resonant (i.e. quasielastic) scattering activity operative on the $k_F$ saddles associated with fermion-to-boson inter-conversion will not contribute towards restricting the optical conductivity. The large quasi-elastic scattering term in evidence has previously been commented upon by Abrahams and Varma (2000ab). They deemed that out-of-plane 'impurities' had to be responsible. However our view of events is even more 'intrinsic' than this – the dopant ions. Besides the above sizeable elastic (and strongly angular dependent) term in the scattering, there exists too of course the very substantial inelastic scattering activity between carriers and between carriers and the lattice. These effects go forward to produce those strong and very characteristic resistivity terms, approximately linear both in $T$ and $\omega$ (Timusk and Statt 1999), which Varma early on (1989) referred to in his Marginal Fermi Liquid formulation of HTSC behaviour as issuing from a 'magic polarizability'. This carrier scattering is at least an order of magnitude more intense than would occur if the transport properties were simply as covered by a standard Boltzmann treatment in conjunction with the LDA calculated band structure. It is not just the unusual $T$ and $\omega$ dependencies to the observed transport properties which are in question, but, above all, the sourcing of their strikingly large prefactors. The situation for the d.c. conductivity long has been recognized, but recently Varma and Abrahams (2001) have offered additional formal treatment within the framework of MFL of how both the Hall and magneto-resistance signals emerge as being likewise dominated by the novel scattering. Through a small angle scattering treatment they endeavour to secure the very characteristic HTSC results $\cot\theta_H$ ($\propto \mu_H^{-1}$) $\propto T^2$ and $\Delta\rho_B \propto T^{-4}$. However, as they now acknowledge, this is not the right way to reach these key relations (Varma and Abrahams 2002; see Carter and Schofield 2002). As stressed in Wilson and Zahrir (1997), the latter are not *in toto* Fermiology results, for they constitute clear expression of that intense fermion-to-boson activity which precipitates HTSC within the present resonant negative-$U$ scenario. Clearly the two-particle, two-subsystem, umklapp and DOS pseudogap aspects to the problem all need to be incorporated explicitly.

Let us now take stock of the present theoretical situation. The magic polarizability looked to within the MFL theory, with its rather momentum independent and structureless form both energetically and thermally, is in accord with a resonant interconversion/quantum critical fluctuation such as occurs in the negative-$U$ scenario. The action as we have seen extends to the lattice, with the accommodation of momentum conservation, as quasiparticles transfer between the saddles, the zone corners and the zone centre. The appropriate short wavelength phonons are observed to couple in to the various charge processes. The local pair bosons still are present to some degree above $T_c$ and they form dispersed propagating modes within the pseudogap. Below $T_c$ the local bosons do not pass in their entirety over into the **K** = 0 condensate and excited dispersed modes for 2$e$ and 2$h$ pairings exist, in addition to the sound-wave-like plasma excitation of the condensate itself. The final condensate comes to hold as many pairs as the *electron* count allows for ($n_s > \frac{1}{2}p$), as a somewhat more standard Cooper pairing around the Fermi surface is precipitated, although the effective overall pair coherence length remains very small (with $H_{c2}$ correspondingly high). Because of the boson-fermion degeneracy and interconversion, the condensate interacts very strongly with the quasi-particle dispersion curves to create kinks there in all directions slightly below $E_F$. The pairing action itself proceeds by virtual excitation from the band structural saddles to the top of the open band at the zone corner, a state which *when unoccupied* is of high energy under antibonding/bonding correlations. The doubly-loaded state however becomes greatly relaxed in negative-$U$ fashion by virtue of its association with band closure and the complete restructuring locally of all the valence band energy states. This occurs in response (specifically in the high-valent subsystem) to a termination of $p/d$ bonding/antibonding operativity as regards $d$-electron Cu-O binding – the essential chemistry within the HTSC problem (Wilson 1988, 2000b). It is a matter of chance for these particular square-planar cuprate materials that the $^{10}Cu_{III}^{2-}$ doubly-loaded state happens to fall in close resonance with $E_F$. Such appears not to be the case with the 3D metallic mixed-valent system $La_4BaCu_5O_{13+\delta}$. The given crystal structure of the HTSC cuprates is doubly favourable in that it also introduces a strong zone edge saddle point very close to $E_F$. From the latter it becomes easy to draw off considerable numbers of bosonic pairings of the general type $e(0,\pi) + e(\pi,0) \rightarrow b(\pi,\pi)$, and this without the participation of phonons or spin fluctuations. Phonon participation principally occurs as bosons are returned to the saddles under the negative-$U$ 'relaxation', or in turn pass from the saddles over into the **K** = 0 condensate. The phonons involved in these charge-coupled processes are longitudinal optic in nature. The bosonic mode excitations as plasma oscillations couple to the LO branches, and become strongly registered via $\mathrm{Im}(-1/\varepsilon)$ within the HREELS experiment.

The 40 meV 'resonance peak' recorded in the neutron scattering experiments (Keimer *et al* 1998, Dai *et al* 1999, Fong *et al* 1999, 2000, He *et al* 2002) is not a bosonic mode capable of supplying the 'glue' for HTSC. Indeed it marks the disruptive excitation of the S = 0 superconducting pairs under S = 1 spin flipping, as a pair is taken back from **K** = (0,0) via $(\pi,\pi)$ to its component fermions near $(\pi,0)$ and $(0,\pi)$. The observed *z*-axis component to this excitation in YBCO-123 and BSCCO-2212 is to do not with magnetic interlayer spin coupling but with the appropriate accommodation of the two dissociating fermionic spins within a bilayer unit cell. The paper by Abanov, Chubukov and colleagues (2002), in preprint form (2001) entitled 'What the $(\pi,\pi)$ resonance peak can do', recognized the spin-exciton view of events endorsed there to be quite distinct from that of antiferromagnetic magnon mediation. Notwithstanding this they continued to look still toward the action of Fermi sea spanning to build up the response function $\chi'(\mathbf{q},\omega)$ around $\mathbf{q} = \mathbf{Q} = (\pi,\pi)$ and $\omega(\mathbf{Q}) = \Omega_{res}$. This 'across the body of the zone' physics has to be contrasted with our 'to the corner of the zone' physics. In order to gain the requisite large 'spin-fermion' coupling constant, *g,* and dimensionless fermion self-energy coupling constant, $\lambda$, of roughly ¾ eV and 2 respectively, they demonstrated it necessary that the 'magnetic' coherence length be just $2a_o$. Accordingly the bosonic coupling mode considered by Abanov *et al* is of very highly overdamped spin excitations, and in this contrary, note, to what popularly is often advocated – an antiferromagnetic spin fluctuation mode acting as bosonic mediator in BCS-related fashion. Pointedly the above value for the *spin* coherence length of just $2a_o$ means it is identical to the range of the superconductive coupling in the local pair view (Wilson 2000b, Quintanilla & Gyorffy 2002); a result that exposes the magnetic ascription of the coupling as not being unique. It is self-evident $g$ and $\lambda$ have to be quite large in order to yield HTSC, but the neutron peak has rather little spectral weight or role in events. Clearly the 'hot spots' on the Fermi surface dominate scattering activity in the zone and dictate the novel transport behaviour above $T_c$, besides the *d*-wave form to the superconductive gapping below, but patently this action is not due to Fermi surface nesting. The uncondensed boson modes dispersing upwards from the hot spots, which the energy-resolved STM scattering experiments from Davis and coworkers (Hoffman *et al* 2002, McElroy *et al* 2003) now appear to bear witness to, reassert the local pair character to the novel events current in these materials – as of course above all does the extreme smallness of $\xi_o^{sc}$.

It would to the author seem essential now to try to draw together the many disparate theoretical approaches pursued in the past, including the spin fluctuation and MFL ones, around the negative-*U* and boson-fermion modelling as currently developed in some detail by de Llano, Tolmachev and colleagues (2003, and refs therein), Tchernyshyov and Ren (1997, 1996), Letz and Gooding (1999), Domanski, Ranninger, Romani, Tripodi and coworkers (2001-3) and several others. In accomplishing this it will be decisive to embrace properly the inhomogeneous two-subsystem nature of the mixed-valent cuprates, which the new STM results and related papers now serve so clearly to illustrate.

## Acknowledgments

I would like to dedicate this paper to Prof B L Gyorffy on the occasion of his 65th birthday and for his helping to make my time in Bristol a long, lively and productive one. At this moment of my own retirement I thank too my wife (P.A.W.) for her inestimable help over a still longer period in seeing this paper and its predecessors brought to fruition. Thanks are in addition due to Dr N E Hussey for his comments on the current manuscript.

**Table 1.**

Characteristics of zone boundary phonons in $YBa_2Cu_3O_7$ propagating *in* basal plane perpendicular and parallel to the chains; *i.e.* along the *a* axis (irreducible representations $\Sigma$) and *b* axis (irreducible representations $\Delta$) respectively. Information drawn from Chung *et al* (2003), figs. 1 and 4 and text. (Note small errors exist there in figure 1, with 1e being labelled LO rather than TO, and superfluous oxygen atoms appearing in 1c, 1e and 1f at the chain level).

| figure label [64] | in plane propag$^n$ vector (re chain) | displacements (re propg$^n$ vector : LO vs. TO) | displacements or 'polarization' re chain | dominant vibration type re Cu-O bonds | phonon mode irreducible representation | $\Gamma$ pt. energy (meV) |
|---|---|---|---|---|---|---|
| 1a | x (⊥) | x (// : LO) | ⊥ [Raman $B_{2g}$] | planar stretching | $\Sigma_{1,4}$ | 72 |
| 1b | y (//) | y (// : LO) | // [Raman $B_{3g}$] | planar & chain stretching | $\Delta_{1,4}$ | 66 |
| 1c' | y (//) | x (⊥ : TO) | ⊥ | planar stretching | $\Delta_{2,3}$ | 72 |
| c'' | x (⊥) | y (// : TO) | // | planar stretching | $\Sigma_{2,3}$ | 66 |
| 1d' | y (//) | y (// : LO) | // [Raman inactive] | chain stretching | $\Delta$ | * |
| d'' | x (⊥) | y (⊥ : TO) | // [Raman inactive] | chain stretching | $\Sigma$ | * |
| 1e' | x (⊥) | z (⊥ : TO) | ⊥ [Raman active] | apical stretching | $\Sigma_1,\Delta_1$ | 62.5 |
| e'' | y (//) | z (⊥ : TO) | ⊥ [IR active] | apical stretching | $\Sigma_4,\Delta_4$ | 69 |
| 1f' | x (⊥) | x (// : LO) | ⊥ | bending of both chain & plane. | $\Sigma$ | 43 |
| f'' | y (//) | x (// : TO) | ⊥ | Ditto | $\Delta$ | * |

LO : longitudinal optic, *i.e*. displacement vector // propagation vector, with Cu and O displacements in bonds in antiphase.

$\Sigma$ vs $\Delta$ splitting is from presence of chains destroying tetragonal symmetry.

- the $\Sigma$ representation wavevectors are //*a* (*i.e*. ⊥ to chains), while the $\Delta$ are //*b* (*i.e*. // to chains).

$\Sigma_1$ and $\Sigma_4$ relate to there being two basal planes per YBCO unit cell. (Likewise $\Delta_2$ and $\Delta_3$).

- of which only ungerade (antisymmetric) inter-layer combination generally seen for integer $l$.

.

**Figures**

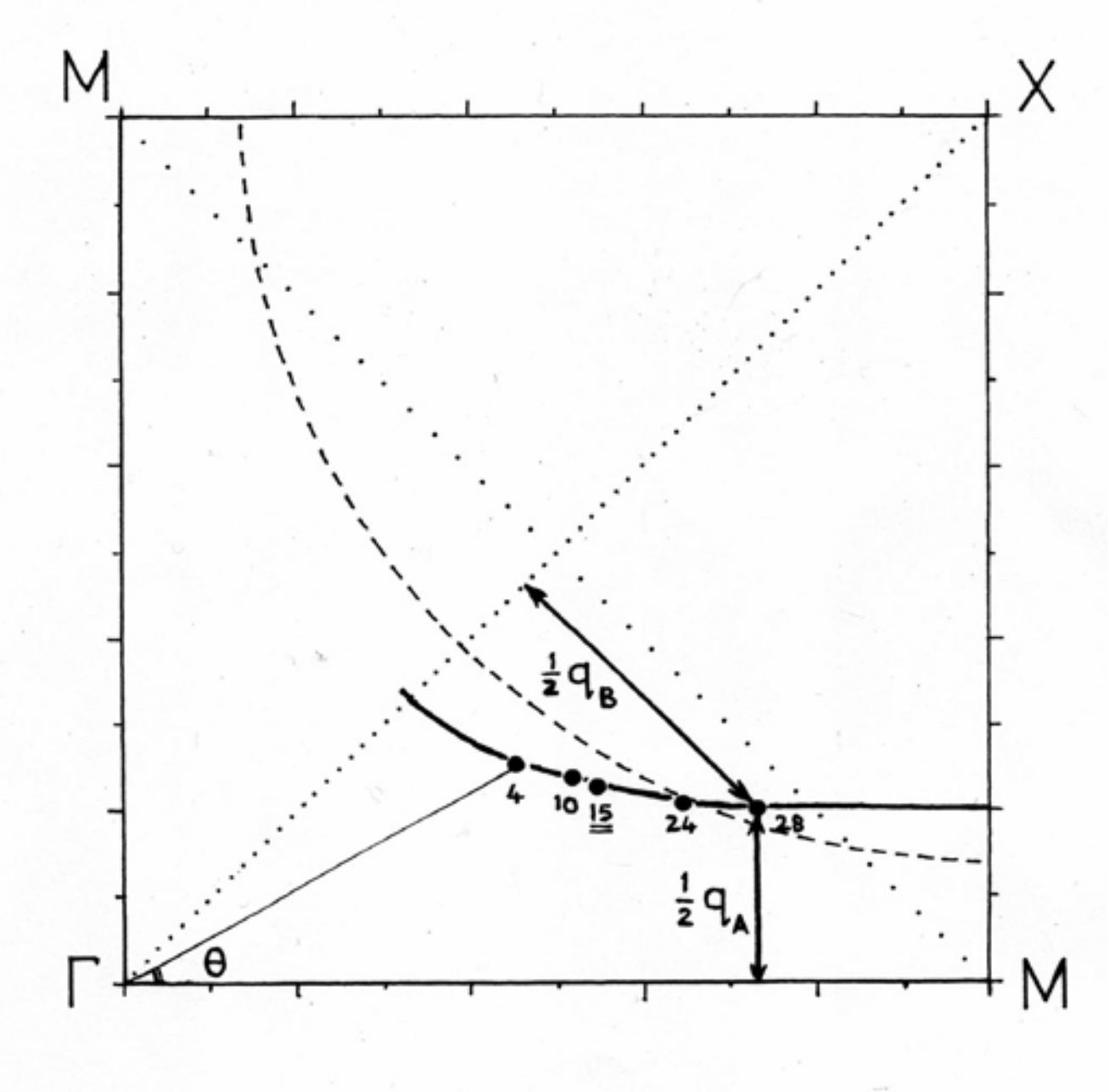

**Figure 1.** Shown in a quadrant of the Brillouin zone is the locus defined by the coupled elastic scattering wavevectors $\mathbf{q}_A$ and $\mathbf{q}_B$ (see text) extracted from Fourier analysis of the STM conductance real space mapping secured at 4 K by Hoffman *et al* [1] for the case of slightly underdoped ($T_c$ = 78 K) $Bi_2Sr_2CaCu_2O_{8+\delta}$. Also shown is the hypothetical circular 2D Fermi surface that would correspond to a hole count relative to half filling (the dashed line) of 0.16, as for optimaly doped HTSC material. The experimental locus obtained here is very similar to that generated in LDA band structural work for the bilayer structured material at optimal doping, as it relates to the fuller of the two sheets to derive from the two Cu-O planes per unit cell (i.e. the bonding interlayer combination). The 'hot spots' are where the M-M dotted line intersects the Fermi surface on the band structural saddles near the M points ('π,0', *etc.*), and the superconducting $d_{x2-y2}$ node lies on the line from 0,0 to 'π,π', close here to one third of the way from zone centre to zone corner. Angular θ values from the $k_x$-axis read off from this plot for each data point are subsequently used in constructing figure 2.

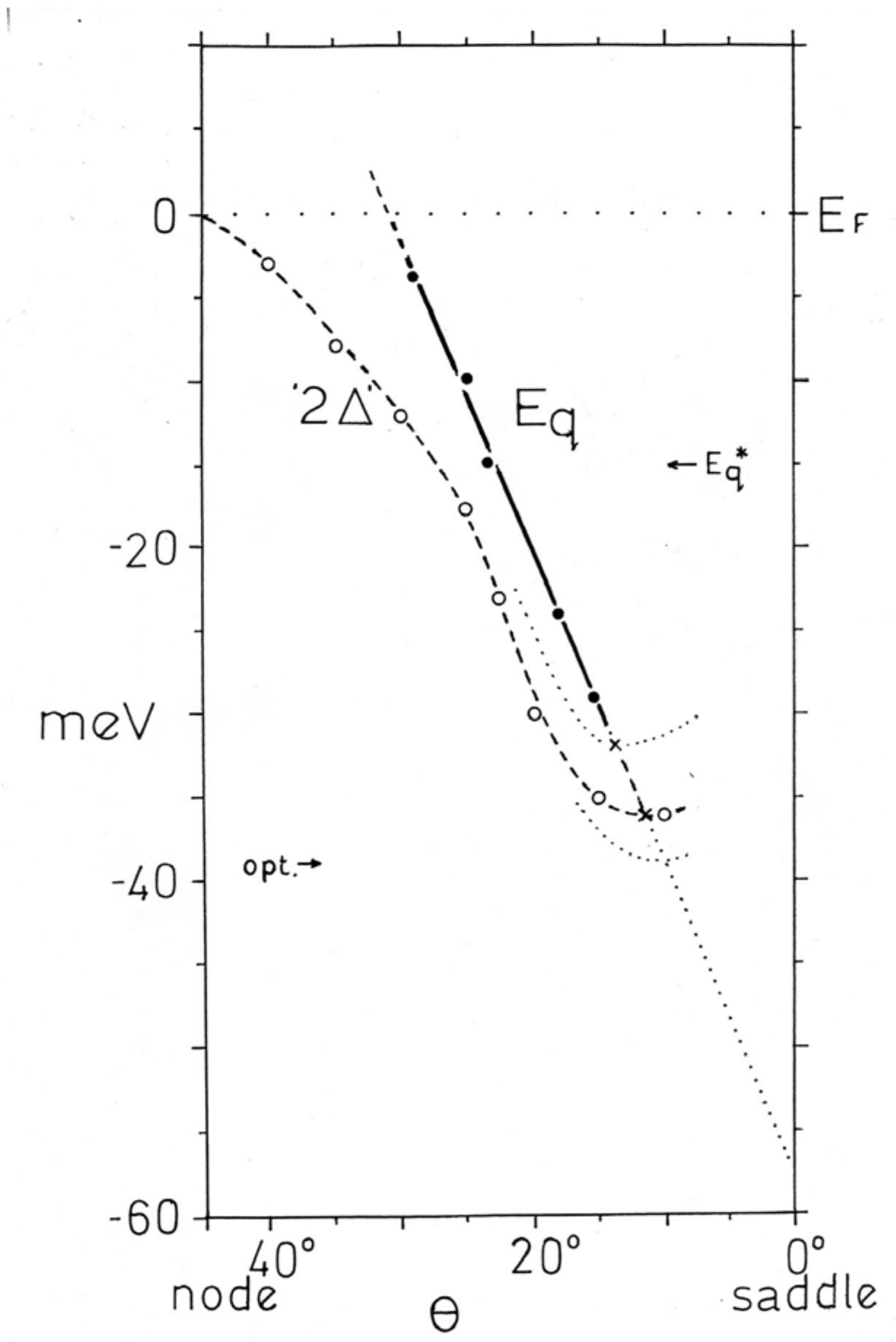

**Figure 2.** The angular variation within the Brillouin zone octant from the nodal to the saddle directions is plotted for the binding energies $E_\mathbf{q}$ (at 4 K) corresponding to the conductance data points located at coupled wavevectors $\mathbf{q}_A$ and $\mathbf{q}_B$ in figure 1. The more or less linear dispersion presented is that for a somewhat underdoped sample. The line appears to shift slowly upwards with a rise in sample hole doping content. The diffuse STM scattering signal is steadily lost as on the one hand the experimental conditions approach the Fermi energy and on the other hand are pushed back down towards the 'hot spots' on the saddles. In the present case the sharpest scattering peak signal was registered for the intermediate condition $\theta \approx 23^o$. The above angular behaviour is to be contrasted with that included in the figure of the first sharp peak in the ARPES data under comparable conditions, this constructed from the data of Mesot *et al* [24]. The latter curved plot (dashed) is taken to be a moderately close guide to the functional form of the superconducting gap $2\Delta(\theta)$ – not $\Delta(\theta)$ as stated in [24]. It is not justified, moreover, to take directly the non-sinusoidal form of this plot as representing faithfully any true deviation from a strictly $d_{x2-y2}$ form for the superconducting order parameter.

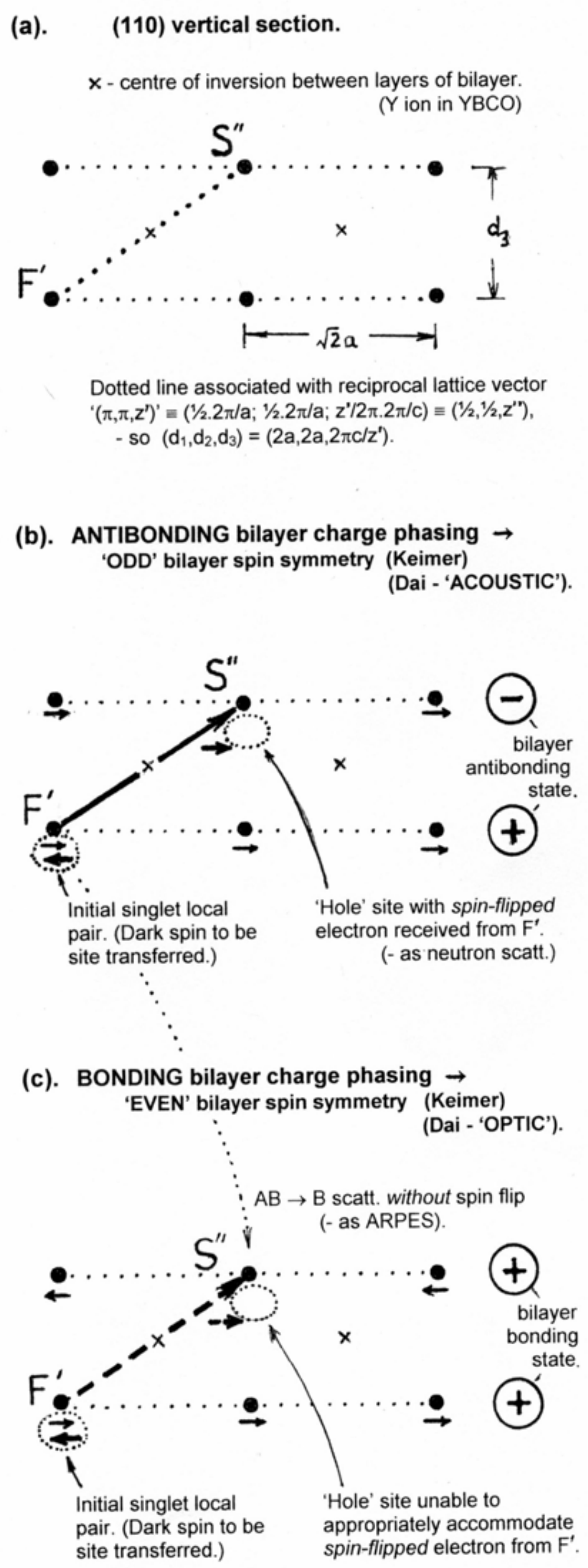

**Figure 3.** The symmetry conditions governing the way in which the bosonic pairs decompose under inelastic neutron spin-flip scattering within the bilayer structures of $YBa_2Cu_3O_7$ and $Bi_2Sr_2CaCu_2O_8$. The situation is shown in vertical (110) section. The appropriate process picks up the $k_z$ component $\zeta' = 2\pi c/d_3$, where $d_3$ is the spacing between Cu-O planes in the bilayer, and it occurs in the antibonding bilayer charge phasing channel.